\documentstyle [12pt]{article}                                                
\begin{document}                                                              
\title{ The Volume Element of Space-Time and Scale Invariance                                  
\protect\\  } \author{E.I. Guendelman                                         
\\{\it Physics Department, Ben-Gurion University, Beer-Sheva                  
84105, Israel}}                                                               
                                                                              
\maketitle                                                                    
\bigskip                                                                      
\begin{abstract}                                                              
Scale invariance is considered  in          
the context of gravitational theories where                                   
the action, in the first order formalism, is of the form $S =                 
\int L_{1} \Phi d^4x$ + $\int L_{2}\sqrt{-g}d^4x$ where the volume 
element  $\Phi d^4x$ is independent of the metric.            
For global scale invariance, a "dilaton"                                      
$\phi$ has to be introduced, with non-trivial potentials $V(\phi)$ =          
$f_{1}e^{\alpha\phi}$ in $L_1$ and $U(\phi)$ = $f_{2}e^{2\alpha\phi}$ in      
$L_2$. This leads to non-trivial mass generation and a potential for          
$\phi$ which is interesting for inflation.
Interpolating models for natural 
transition from inflation to a slowly accelerated universe at late
times appear naturally. This is also achieved for "Quintessential models",
which are scale invariant but formulated with the use of volume 
element  $\Phi d^4x$ alone.
For closed strings and branes (including the supersymmetric cases),
the modified measure formulation is possible and does not require 
the introduction of a particular scale (the string or brane tension) from the
begining but rather these appear as integration constants.

\end{abstract}
\section{The Simplest Scalar-Gravity Model, in the absence of fermions}

The concept of scale invariance appears as an attractive possibility for a    
fundamental symmetry of nature. In its most naive realizations, such a        
symmetry is not a viable symmetry, however, since nature seems to have        
chosen some typical scales.                                                   
                                                                              
Here we will find that scale invariance can nevertheless be incorporated      
into realistic, generally covariant field theories. However, scale            
invariance has to be discussed in a more general framework than that of       
standard generally relativistic theories, where we must allow in the          
action, in addition to the                                                    
ordinary measure of integration $\sqrt{-g}d^{4}x$, another one,               
$\Phi d^{4}x$, where $\Phi$ is a density built out of degrees of freedom      
independent of the metric.                                                    
                                                                              
        For example, given 4-scalars $\varphi_{a}$ (a =                      
1,2,3,4), one can construct the density                                       
\begin{equation}                                                              
\Phi =  \varepsilon^{\mu\nu\alpha\beta}  \varepsilon_{abcd}                   
\partial_{\mu} \varphi_{a} \partial_{\nu} \varphi_{b} \partial_{\alpha}       
\varphi_{c} \partial_{\beta} \varphi_{d}                                      
\end{equation}                                                                
                                                                              
        One can allow both geometrical                                        
objects in the theory and consider $^1$ ,                                  
\begin{equation}                                                              
S = \int L_{1} \Phi  d^{4} x  +  \int L_{2} \sqrt{-g}d^{4}x                   
\end{equation}                                                                
                                                                              
         Here $L_{1}$ and $L_{2}$ are                                         
$\varphi_{a}$  independent. There is a good reason not to consider            
mixing of  $\Phi$ and                                                         
$\sqrt{-g}$ , like                                                            
for example using                                                             
$\frac{\Phi^{2}}{\sqrt{-g}}$. This is because (2) is invariant (up to the inte
divergence) under the infinite dimensional symmetry                           
$\varphi_{a} \rightarrow \varphi_{a}  +  f_{a} (L_{1})$                       
where $f_{a} (L_{1})$ is an arbitrary function of $L_{1}$ if $L_{1}$ and      
$L_{2}$ are $\varphi_{a}$                                                     
independent. Such symmetry (up to the integral of a total divergence) is      
absent if mixed terms are present.                                            
                                                                              
        We will study now the dynamics of a scalar field $\phi$ interacting   
with gravity as given by the action (2) with $^{2,3,4}$                              
\begin{equation}                                                              
L_{1} = \frac{-1}{\kappa} R(\Gamma, g) + \frac{1}{2} g^{\mu\nu}               
\partial_{\mu} \phi \partial_{\nu} \phi - V(\phi),  L_{2} = U(\phi)           
\end{equation}                                                                
                                                                              
\begin{equation}                                                              
R(\Gamma,g) =  g^{\mu\nu}  R_{\mu\nu} (\Gamma) , R_{\mu\nu}                   
(\Gamma) = R^{\lambda}_{\mu\nu\lambda}, R^{\lambda}_{\mu\nu\sigma} (\Gamma) = 
\Gamma^{\lambda}_{\mu\nu,\sigma} - \Gamma^{\lambda}_{\mu\sigma,\nu} +                          
\Gamma^{\lambda}_{\alpha\sigma}  \Gamma^{\alpha}_{\mu\nu} -                   
\Gamma^{\lambda}_{\alpha\nu} \Gamma^{\alpha}_{\mu\sigma}.                     
\end{equation}                                                                
                                                                              
        In the variational principle $\Gamma^{\lambda}_{\mu\nu},              
g_{\mu\nu}$, the measure fields scalars                                       
$\varphi_{a}$ and the  scalar field $\phi$ are all to be treated              
as independent variables.                                                     
        If we perform the global scale transformation ($\theta$ =             
constant)                                                                     
\begin{equation}                                                              
g_{\mu\nu}  \rightarrow   e^{\theta}  g_{\mu\nu}                              
\end{equation}                                                                
then (2), with the definitions (3), (4), is invariant provided  $V(\phi)$     
and $U(\phi)$ are of the                                                      
form                                                                          
\begin{equation}                                                              
V(\phi) = f_{1}  e^{\alpha\phi},  U(\phi) =  f_{2}                            
e^{2\alpha\phi}                                                               
\end{equation}                                                                
and $\varphi_{a}$ is transformed according to                                 
$\varphi_{a}   \rightarrow   \lambda_{a} \varphi_{a}$                         
(no sum on a) which means 
$\Phi \rightarrow \biggl(\prod_{a} {\lambda}_{a}\biggr) \Phi \equiv \lambda   
\Phi $                                                                        
such that                                                                     
$\lambda = e^{\theta}$                                                        
and                                                                           
$\phi \rightarrow \phi - \frac{\theta}{\alpha}$. In this case we call the     
scalar field $\phi$ needed to implement scale invariance "dilaton".

\subsection{Equations of Motion}
        Let us consider the equations which are obtained from                 
the variation of the $\varphi_{a}$                                            
fields. We obtain then  $A^{\mu}_{a} \partial_{\mu} L_{1} = 0$                
where  $A^{\mu}_{a} = \varepsilon^{\mu\nu\alpha\beta}                         
\varepsilon_{abcd} \partial_{\nu} \varphi_{b} \partial_{\alpha}               
\varphi_{c} \partial_{\beta} \varphi_{d}$. Since                              
det $(A^{\mu}_{a}) =\frac{4^{-4}}{4!} \Phi^{3} \neq 0$ if $\Phi\neq 0$.       
Therefore if $\Phi\neq 0$ we obtain that $\partial_{\mu} L_{1} = 0$,          
 or that                                                                      
$L_{1}  = M$,                                                                 
where M is constant. This constant M appears in a self-consistency            
condition of the equations of motion                                          
that allows us to solve for $ \chi \equiv \frac{\Phi}{\sqrt{-g}}$             
                                                                              
\begin{equation}                                                                 
\chi = \frac{2U(\phi)}{M+V(\phi)}.                                            
\end{equation}                                                                
                                                                              
        To get the physical content of the theory, it is convenient to go     
to the Einstein conformal frame where                                         
\begin{equation}                                                              
\overline{g}_{\mu\nu} = \chi g_{\mu\nu}                                       
\end{equation}                                                                
and $\chi$  given by (7). In terms of $\overline{g}_{\mu\nu}$   the non       
Riemannian contribution (defined   as                                         
$\Sigma^{\lambda}_{\mu\nu} =                                                  
\Gamma^{\lambda}_{\mu\nu} -\{^{\lambda}_{\mu\nu}\}$                           
where $\{^{\lambda}_{\mu\nu}\}$   is the Christoffel symbol),                 
disappears from the equations, which can be written then in the Einstein      
form ($R_{\mu\nu} (\overline{g}_{\alpha\beta})$ =  usual Ricci tensor)        
\begin{equation}                                                              
R_{\mu\nu} (\overline{g}_{\alpha\beta}) - \frac{1}{2}                         
\overline{g}_{\mu\nu}                                                         
R(\overline{g}_{\alpha\beta}) = \frac{\kappa}{2} T^{eff}_{\mu\nu}             
(\phi)                                                                        
\end{equation}                                                                
where                                                                         
\begin{equation}                                                              
T^{eff}_{\mu\nu} (\phi) = \phi_{,\mu} \phi_{,\nu} - \frac{1}{2} \overline     
{g}_{\mu\nu} \phi_{,\alpha} \phi_{,\beta} \overline{g}^{\alpha\beta}          
+ \overline{g}_{\mu\nu} V_{eff} (\phi),                                       
V_{eff} (\phi) = \frac{1}{4U(\phi)}  (V+M)^{2}.                               
\end{equation}                                                                
        If $V(\phi) = f_{1} e^{\alpha\phi}$  and  $U(\phi) = f_{2}
e^{2\alpha\phi}$ as                                                           
required by scale invariance, we obtain from (10)                             
\begin{equation}                                                              
        V_{eff}  = \frac{1}{4f_{2}}  (f_{1}  +  M e^{-\alpha\phi})^{2}        
\end{equation}                                                                
                                                                              
        Since we can always perform the transformation $\phi \rightarrow      
- \phi$ we can                                                                
choose by convention $\alpha > O$. We then see that as $\phi \rightarrow      
\infty, V_{eff} \rightarrow \frac{f_{1}^{2}}{4f_{2}} =$ const.                
providing an infinite flat region. Also a minimum is achieved at zero         
cosmological constant , without fine tuning 
for the case $\frac{f_{1}}{M} < O$ at the point         
$\phi_{min}  =  \frac{-1}{\alpha} ln \mid\frac{f_1}{M}\mid $. Finally,        
the second derivative of the potential  $V_{eff}$  at the minimum is          
$V^{\prime\prime}_{eff} = \frac{\alpha^2}{2f_2} \mid{f_1}\mid^{2} > O$        

\section{ Some Physics of the Model: Inflation, Connection to Zee's 
Induced Gravity Model and Possible applications to the Present state of the
Universe }

        A very important  point to be raised is that since there is an 
infinite      
region of flat potential for $\phi \rightarrow \infty$, we expect a slow      
rolling                                                                    
inflationary scenario to be                                               
viable, provided the universe is started at a sufficiently large value of     
the scalar field $\phi$ for example.               
                
The fact that there is a flat region is directly correlated to the fact that
there is scale invariance. In fact, in terms of 
$\overline{g}_{\mu\nu} $ and $\phi$, the scale transformations affect only
$\phi$ ($\overline{g}_{\mu\nu}$ is scale invariant) and it is simply a translation
in the scalar field space. The flat region reflects
 a translation invariant region, where therefore scale invariance is 
restored. By contrast any non
trivial shape of the potential  means ssb
of scale invariance, as is the case in a region of the potential 
(for $M \neq 0$).

It is also very interesting to notice that the theory can be related to the
induced gravity theory of Zee $^5$, defined by the action,                              
                                                                                                              
\begin{equation}                                                              
S  = \int \sqrt{-g} (-  \frac {1}{2} \epsilon \varphi^{2} R +
\frac{1}{2} g^{\mu\nu} \partial_{\mu} \varphi \partial_{\nu} \varphi
-\frac {\lambda}{8} ( \varphi^{2} - \eta^{2} )^{2} ) d^{4} x                    
\end{equation}                                                                
                                                                              
Here it is assumed that the second order formalism is used, i.e.              
$R=R(g)=$ usual Riemannian scalar curvature defined in terms of $ g_{\mu\nu}$.
Notice that if  $ \eta = 0 $, the action is        
invariant under the global scale transformation                               
$g_{\mu\nu}  \rightarrow   e^{\theta}  g_{\mu\nu}$, $\varphi \rightarrow
e^{-\frac{\theta}{2}} \varphi$, but a finite induced Newton's constant
is defined only if $ \eta $ is non vanishing. Then defining ($2k^{2} =\kappa$) 
$\overline{g}_{\mu\nu} = 
k^{2} \epsilon \varphi^{2} g_{\mu\nu}$ and the scalar field                                                                                                 
$ \phi = \frac {1}{k} \sqrt{ 6 + \frac {1}{\epsilon} } ln \varphi $,                            
one can then show that the induced gravity  model is equivalent to standard    
General  Relativity (expressed in terms of $\overline{g}_{\mu\nu}$ )
 minimally coupled to the scalar field $\phi$ which has a potential $^6$ , 
$V_{eff}  = \frac{\lambda}{8 k^{4} \epsilon^{2}}  (1  - 
 \eta^{2} e^{-2\sqrt{ \frac {\epsilon}{1 + 6 \epsilon} } k \phi})^{2}$                                                            
 which is exactly the form (11) with 
$\alpha = 2\sqrt{ \frac {\epsilon}{1 + 6 \epsilon}} k $ (in Ref.6, $k$ is called
$\kappa$). 
The induced gravity model (12) is quite 
successful from the point of view of its applications to inflation and it 
has been studied by a number of authors in this context $^7$ . 
Notice that the induced gravity model is not 
consistent with scale invariance for a non vanishing $\eta$, while the theory 
developed here, which leads to the induced gravity model after ssb, 
has been constructed
starting with scale invariance as a fundamental principle.

A similar thing happens when we take the pure gravity form 
(see Refs. 8, 9 and 10),                        
$S  = \frac {1}{2}\int \sqrt{-g} (R + \beta R^{2} ) d^{4} x $           
Here again it is assumed that the second order formalism is used, i.e.   
$R=R(g)=$ usual Riemannian scalar curvature defined in terms of $ g_{\mu\nu}$.
Notice that if only the $ \beta R^{2} $ term is present, the action is        
invariant under the global scale transformation                               
$g_{\mu\nu}  \rightarrow   e^{\theta}  g_{\mu\nu}$.

Then defining  $\overline{g}_{\mu\nu} = [1 + 2\beta R(g)] g_{\mu\nu}$                          
 and $\phi = \sqrt{ \frac {3}{2} } ln [1 + 2\beta R(g)]$,
 one can then show that the $ \beta R^{2} $ model is equivalent 
to standard    
General  Relativity (expressed in terms of $\overline{g}_{\mu\nu}$ ) 
coupled to
a minimally coupled scalar field which has a potential (see for example 
Ref. 9 )
$V_{eff}  = \frac{1}{8 \beta}  (1  -   e^{-\sqrt{ \frac {2}{3} }\phi})^{2}$,
which is exactly the form (11) for a very special choice of  $\alpha$         
($ =\sqrt{ \frac {2}{3}}$ in Planck units).

Notice that as $R^{2}$ dominates, $\phi \rightarrow \phi + const. $  under a  
dilatation transformation and one can again understand the flat region as a   
consequence of scale invariance in some limit.

Density fluctuations have been studied in the $ \beta R^{2} $ model $^{10}$
 and it was found that 
$10^{11} Gev < \sqrt { \frac {1}{ \beta} } < 10^{13} Gev$
gives acceptable density fluctuations. 
Notice that when changing continously the parameters in the Zee model, 
we obtain a correspondence with with the theory defined here for a 
continous range of the $\alpha$ parameter, while for the 
$ \beta R^{2} $ the correspondence is 
achieved for a very specific value of $\alpha$ only.

Furthermore, independently of the question of what kind of models can be 
equivalent (before we couple it to matter) to the scale invariant theory
defined here, one can consider this model as suitable for the      
present day universe rather than for the early universe, after we suitably    
reinterpret the meaning of the scalar field  $\phi$. This can provide a long  
lived almost constant vacuum energy for a                                     
long period of time, which can be small if $f_{1}^{2}/4f_{2}$ is              
small. Such small energy                                                      
density will eventually disappear when the universe achieves its true         
vacuum state.                                                                 
                                                                              
        Notice that a small value of $\frac{f_{1}^{2}}{f_{2}}$   can be       
achieved if we let $f_{2} >> f_{1}$. In this case                             
$\frac{f_{1}^{2}}{f_{2}} << f_{1}$, i.e. a very small scale for the           
energy                                                                        
density of the universe is obtained by the existence of a very high scale     
(that of $f_{2}$) the same way as a small fermion mass is obtained in the     
see-saw mechanism $^{11}$ from the existence also of a large mass scale. 

\section{The Introduction of Fermions}
    
Since in nature there is more than just scalars and gravity, it is necessary
to consider the extension of the model so as to accomodate fermions.
Taking, for example, the case of a fermion $\psi$, where the kinetic term     
of the fermion is chosen to be part of $L_1$                                  
\begin{equation}                                                              
S_{fk} = \int L_{fk} \Phi d^4 x                                               
\end{equation}                                                                
\begin{equation}                                                              
L_{fk} = \frac{i}{2} \overline{\psi} [\gamma^a V_a^\mu                        
(\overrightarrow{\partial}_\mu + \frac{1}{2} \omega_\mu^{cd} \sigma_{cd})     
- (\overleftarrow{\partial}_\mu + \frac{1}{2} \omega_\mu^{cd} \sigma_{cd})    
\gamma^a V^\mu_a] \psi                                                        
\end{equation}                                                                
there $V^\mu_a$ is the vierbein, $\sigma_{cd}$ =                              
$\frac{1}{2}[\gamma_c,\gamma_d]$, the spin connection $\omega^{cd}_\mu$ is    
determined by variation with respect to $\omega^{cd}_\mu$ and, for            
self-consistency, the curvature scalar is taken to be (if we want to deal     
with $\omega_\mu^{ab}$ instead of $\Gamma^\lambda_{\mu\nu}$ everywhere)       
\begin{equation}                                                              
R = V^{a\mu}V^{b\nu}R_{\mu\nu ab}(\omega),                                    
R_{\mu\nu ab}(\omega)=\partial_{\mu}\omega_{\nu ab}                           
-\partial_{\nu}\omega_{\mu ab}+(\omega_{\mu a}^{c}\omega_{\nu cb}             
-\omega_{\nu a}^{c}\omega_{\mu cb}).                                          
\end{equation}                                                                
                                                                              
Global scale invariance is obtained                                           
provided $\psi$ also transforms, as in                                        
$\psi \rightarrow \lambda ^{-\frac{1}{4}} \psi$. Mass term consistent with    
scale invariance exist,                                                       
\begin{equation}                                                              
S_{fm} = m_1 \int \overline{\psi} \psi e^{\alpha\phi/2} \Phi d^4x + m_2       
\int \overline{\psi} \psi e^{3\alpha\phi/2} \sqrt{-g} d^4 x.                  
\end{equation}

If we consider the situation where                                            
$m_1 e^{\alpha\phi/2} \overline{\psi}\psi$                                    
or $m_2 e^{3\alpha\phi/2} \overline{\psi}\psi$ are                            
much bigger than $V(\phi)$ + M, i.e. a high density approximation, we         
obtain that instead of (7) that the consistency condition is $^3$              
 $(3m_2 e^{3\alpha\phi/2} + m_1 e ^{\alpha\phi/2} \chi) \overline{\psi}       
\psi = 0$,                                                                    
which means                                                                   
$\chi = -\frac{3m_2}{m_1} e^{\alpha\phi}$. Using this in (16), we obtain,     
after going to the conformal Einstein frame, which involves,                  
also a transformation of the fermion fields,                                  
necessary so as to achieve                                                    
Einstein-Cartan form for both the gravitational and fermion
equations. These transformations are,                                         
$\overline{g}_{\mu\nu}$ = $\chi g_{\mu\nu}$ (or                               
$\overline V_\mu^a$ = $\chi^\frac{1}{2} V_\mu^a$) and $\psi ^\prime$ =        
$\chi ^{-\frac{1}{4}} \psi$ and they lead to a mass term,                     
\begin{equation}                                                              
S_{fm} = -2m_2 ( \frac {|m_1|}{3|m_2|})^{3/2}  \int\sqrt{-\overline {g}}      
\overline{\psi} ^{\prime} \psi ^{\prime} d^4x                                 
\end{equation}                                                                
                                                                              
The $\phi$ dependence of the mass term has disappeared, i.e. masses are       
constants.                                                                    
                                                                              
There is one situation where the low density of matter can also give results  
which are similar to those obtained in the high density approximation, in     
that the coupling of the $ \phi $ field disappears and that the mass term
becomes of a conventional form in the Einstein conformal frame.               
                                                                              
This is the case, when we study the theory for the limit                      
$\phi \rightarrow \infty$ . Then $U(\phi) \rightarrow \infty$ and             
$V(\phi) \rightarrow \infty$. In this case, taking                            
$m_1 e^{\alpha\phi/2} \overline{\psi}\psi$                                    
and $m_2 e^{3\alpha\phi/2} \overline{\psi}\psi$                               
much smaller than $V(\phi)$ or $U(\phi)$ respectively, therefore one can see  
that (7) is a good approximation and since                                    
also $M$ can be ignored in the self consistency condition (7)                 
in this limit, we get then,                                                   
$\chi = \frac{2f_2}{f_1} e^{\alpha\phi}$. If this is inserted in (16),        
we get $S_{fm} = m \int\sqrt{-\overline {g}}                                  
\overline{\psi} ^{\prime} \psi ^{\prime} d^4x$, where                         
                                                                              
\begin{equation}                                                              
 m = m_1(\frac {f_1}{2f_2})^{\frac{1}{2}} +  
m_2(\frac {f_1}{2f_2})^{\frac{3}{2}}
\end{equation}                                                                
                                                                              
Comparing (17) and (18) and taking  $m_1$ and $m_2$ of the same               
order of magnitude, we see that the mass of the Dirac particle is much        
smaller in the region $\phi \rightarrow \infty$, for which (18) is            
valid, than it is in the region of high density of the Dirac particle         
relative to $V(\phi)+M$, as displayed in eq. (17), if the "see-saw"           
assumption $\frac{f_1}{f_2} < < 1$ is made. Therefore if 
space is populated by these diluted Dirac particles of this
type, the mass of these particles will grow substantially if we 
go to the true vacuum state, valid in the absence of matter, 
i.e. $V+M=0$, as dictated by $V_{eff}$ given by eq. (11).
                                                                              
The presence of matter pushes therefore the minimum of energy to a state      
where $ V+M > 0$. The real vacuum in the presence of matter should not        
be located in the region $\phi \rightarrow \infty$, which minimizes the       
matter energy, but maximizes the potential energy $V_{eff}$ and not at        
$V+M=0$, which minimizes $V_{eff}$, and where particle masses are big, but    
somewhere in a balanced intermediate stage. Clearly how much above $V+M=0$    
such true vacuum is located must be correlated to how much particle density   
is there in the Universe. A non zero vacuum energy, which must be of the      
same order of the particle energy density, has to appear and this could       
explain the "accelerated universe" that appears to be implied by the most     
recent observations, together with the "cosmic coincidence", that requires    
the vacuum energy be of the same order of magnitude to 
the matter energy $^{12}$ . 
                                                  
\section{On The Absence of the Goldstone Boson}                             
It is worthwhile to point out that in the models with scale                   
invariance dicussed here there is no Goldstone boson, when we look at the     
excitations arround the true vacuum with zero cosmological constant. The      
basic reason that Goldstone's theorem does not apply is that although         
there is a global symmetry, which leads, according to Noether's theorem       
to a locally conserved current, the spatial components of such current have   
an infrared singular behavior, leading to flux leaking through infinity and   
to a non conservation of the would be dilaton charge $^4$ .                  

Let us see that this is indeed the case and for this purpose,                 
let us ignore the fermions. Since there is a symmetry                         
according to  Noether's theorem, there                                        
is a conserved current given by (since the variation of the lagrangian        
density vanishes under the scale symmetry),                                   
                                                                              
\begin{equation}                                                              
j^{\mu} = \frac {\partial L} {\partial (\partial_{\mu} \varphi_{a})} \delta \varphi_{a}
+ \frac {\partial L} {\partial (\partial_{\mu} \phi)} \delta \phi             
\end{equation}                                                                
                                                                              
since in the first order formalism                                            
$ \frac {\partial L} {\partial (\partial_{\mu} g_{\alpha \beta})} = 0 $ and   
$\delta \Gamma^{\lambda}_{\mu\nu}= 0 $ under the                              
scale symmetry defined before.

Let us now consider what we should take for $\delta \varphi_{a}$. 
As part of the dilatation symmetry, we have that                              
$\varphi_{a} \rightarrow \lambda_{a} \varphi_{a}$ (no sum on a) and since     
$\biggl(\prod_{a} {\lambda}_{a}\biggr) \equiv \lambda= e^{\theta}$, we have,  
taking a transformation infinitesimally close to the identity, i.e.           
${\lambda}_{a} = 1 + {\epsilon}_{a}$, with ${\epsilon}_{a} << 1$ and          
all ${\epsilon}_{a}$ equal, so that ${\epsilon}_{a} = {\theta}/4$ and since   
also $\delta \phi = - \frac {\theta} {\alpha} $, that the conserved dilatation
current is,                                                                   
\begin{equation}                                                              
j^{\mu}_{\theta} = -\frac {\theta}{\alpha} \Phi \partial^{\mu} \phi +         
\theta \varepsilon^{\mu\nu\alpha\beta} \varepsilon_{abcd} \varphi_{a}         
\partial_{\nu} \varphi_{b} \partial_{\alpha}                                  
\varphi_{c} \partial_{\beta} \varphi_{d} L_{1}  \equiv \theta j^{\mu}_{D}     
\end{equation}                                                                

To see the basic reasons why the dilatation current  has an infrared          
singular behavior, let us consider the spatial behavior of the                
$\varphi_{a}$ fields for the case of a simple spatially flat Robertson-Walker 
solution of the form                                                          
\begin{equation}                                                              
ds^{2} = -dt^{2} + R^{2}(t) (dx^{2} +dy^{2} + dz^{2}),                        
 \phi = \phi(t)                                                               
\end{equation}                                                                
 We see also from the constraint (7) that $\chi = \chi(t)$. Then, since       
$\chi = \chi(t) = \frac {\Phi}{R^{3}(t)}$, we get that,                       
\begin{equation}                                                              
\Phi = R^{3}(t) \chi(t) = \varepsilon^{\mu\nu\alpha\beta}  \varepsilon_{abcd} 
\partial_{\mu} \varphi_{a} \partial_{\nu} \varphi_{b} \partial_{\alpha}       
\varphi_{c} \partial_{\beta} \varphi_{d}                                      
\end{equation}                                                                

This can be solved by taking                                                  
\begin{equation}                                                              
\varphi_{1} = x, \varphi_{2} = y, \varphi_{3} = z,                            
\varphi_{4} = - \frac{1}{4!} \int \chi(t^{'}) R^{3}(t^{'}) dt^{'}             
\end{equation}                                                                
For this case, with a time dependent scalar field  $\phi(t)$ and with         
$\varphi_{a}$ given above, the spatial components of the current              
$j^{\mu}_{D}$, diverge linearly as $x^{i} \rightarrow \infty$                 
($x^{1}=x, x^{2}=y, x^{3}=z$). In fact                                        
$j^{i}_{D} \rightarrow M x^{i} \chi(t) R^{3}(t)$ as                           
$x^{i} \rightarrow \infty $                                                   
Such current does indeed give flux at infinity. The current grows 
linearly with
distance, so that the total flux is proportional to the volume enclosed       
and obviously the total dilatation charge is not conserved here.

\section {Interpolating Models}
This kind of theories can naturally provide a dynamics that          
interpolates between a high energy density (associated with inflation)        
and a very low energy density (associated with the present universe).         
For this consider two scalar fields $\phi_{1}$ and $\phi_{2}$, 
with normal kinetic terms coupled to
 $\Phi$ as it has been done with the simpler model of just one scalar field.  
Introducing for $\phi_{1}$ a potential $V_{1}(\phi_{1}) =                     
a_{1} e^{\alpha_{1}\phi_{1}}$ that couples to $\Phi$  and another             
$U_{1}(\phi_{1}) = b_{1}e^{2\alpha_{1} \phi_{1}}$                             
 that couples to $\sqrt{-g}$ as required by scale invariance and the potential
for $\phi_{2}$,                                                               
 $V_{2}(\phi_{2}) = a_{2} e^{\alpha_{2}\phi_{2}}$  that couples to $\Phi$  and
 $U_{2}(\phi_{2}) = b_{2}e^{2\alpha_{2}\phi_{2}}$                             
that couples to $\sqrt{-g}$, we arrive (after going through the same steps    
as those explained in the model with just one scalar, i.e. 
solving the constraint and
going to the Einstein frame) at                                               
the effective potential (see the last reference of Ref.3)

\begin{equation}                                                              
        V_{eff}  =                                                            
\frac{(V_{1}(\phi_{1})  + V_{2}(\phi_{2}) + M )^{2}}{4(U_{1}(\phi_{1})  +
 U_{2}(\phi_{2}))}
\end{equation}                                                                
                                                                              
which introduces interactions between $\phi_{1}$  and $\phi_{2}$, although no 
interactions appeared in the original action (i.e. no direct couplings        
appeared). If we take then $\alpha_{1} \phi_{1}$ very big while $\phi_{2}$    
is fixed, then $ V_{eff} $ approaches the constant value                      
$\frac{a_{1}^{2}}{4b_{1}}$ while if we take $\alpha_{2} \phi_{2}$             
to be very big while  $\phi_{1}$  is kept fixed, then $ V_{eff} $             
approaches the constant value                                                 
$\frac{a_{2}^{2}}{4b_{2}}$. One of these flat regions of the potential can 
be associated with a very high energy density, associated with inflation and the 
other can be very small and associated with the energy density of the present 
universe. The effective potential (24) provides therefore a dynamics that     
interpolates naturally between the inflationary phase and the                 
present slowly accelerated universe.                                          

\section{Scale Invariant Quintessential Models} 
One may wonder whether a model that uses only one measure, the measure
 $\Phi$ is possible.
If we follow the most straightforward approach and take the limit in (3) 
$L_{2} = U(\phi) \rightarrow 0 $, or for the scale invariant case
$f_{2} \rightarrow 0$, we see that the potential in (11) forces, in this 
singular limit, the function $f_{1} +Me^{\alpha \phi}$ to vanish, therefore
killing the scalar field dynamics.

It is however possible to restore non trivial scalar field dynamics, while 
keeping the simple structure 

\begin{equation}
S = \int \Phi L d^{4}x
\end{equation}  

which has the invariance $L \rightarrow L + constant $ and therefore shows 
the "principle of non gravitating vacuum energy" (i.e. the irrelevance of
the origin of L) in its most pure form.

The clue to obtain non trivial scalar field dynamics (and also gauge dynamics)
consists in introducing a four index field strength 
$F_{\mu \nu \alpha \beta} = \partial_{[\mu} A_{\nu \alpha \beta]}$, as 
is discussed in Ref.$1$.

Here we will study such kind of models, subjected to the additional 
requirement of scale invariance, which is the unifiying feature of all
the models studied in this paper.

As we will see, such a construction naturally leads to quintessential 
type potentials $^{13}$ which are of interest in cosmology.
   
Let us define the scalar field 

\begin{equation}
y = \frac{1}{m^{2}} \frac {\varepsilon^{\mu\nu\alpha\beta}}{\sqrt{-g}}  
\partial_{\mu} A_{\nu \alpha \beta} 
\end{equation}                

where $m$ is a parameter with the dimensions of a mass. Then let us take
an action with the four field strength and a scalar field $\phi$ according to

\begin{equation} 
L =  \frac{-1}{\kappa} R(\Gamma, g) -\frac{1}{pm^{4(p-1)}}y^{p} + 
\frac{1}{2} g^{\mu\nu} \partial_{\mu} \phi \partial_{\nu} \phi - V(\phi)
\end{equation}  

where the curvature scalar $R(\Gamma, g)$ is once again defined by (4),
$p$ is a dimensionless number and scale invariance requires an 
exponential form for the scalar field potential,

\begin{equation} 
V(\phi) = f e^{\alpha \phi}
\end{equation}

Under these circumtances, S as given by (25), (26), (27) and (28) 
will be invariant under the scale transformations
   
\begin{equation}                                                              
g_{\mu\nu}  \rightarrow   e^{\theta}  g_{\mu\nu},
\phi \rightarrow \phi - \frac{\theta}{\alpha}
\end{equation}
\begin{equation}
\varphi_{a} \rightarrow  \lambda_{a} \varphi_{a} 
\end{equation}  
(no sum on a),      which means 
\begin{equation} 
\Phi \rightarrow \biggl(\prod_{a} {\lambda}_{a}\biggr) \Phi \equiv \lambda   
\Phi, where  \lambda =  e^{\theta}           
\end{equation} 
and finally 
\begin{equation} 
A_{\nu \alpha \beta} \rightarrow   e^{\theta (2-\frac {1}{p})} A_{\nu \alpha \beta}
\end{equation}   
as before the variation with respect to the fields $\varphi_{a}$ gives rise
to the equation  $A^{\mu}_{a} \partial_{\mu} L = 0$,
where  $A^{\mu}_{a} = \varepsilon^{\mu\nu\alpha\beta}
\varepsilon_{abcd} \partial_{\nu} \varphi_{b} \partial_{\alpha}               
\varphi_{c} \partial_{\beta} \varphi_{d}$. Since                              
det $(A^{\mu}_{a}) =\frac{4^{-4}}{4!} \Phi^{3} \neq 0$ if $\Phi\neq 0$.       
Therefore if $\Phi\neq 0$ we obtain that $\partial_{\mu} L = 0$, or that                                                                      

\begin{equation}    
L  = M
\end{equation} 

where $M$ is a constant.
It is clear that if $M \neq 0$, the above equation spontaneously breaks 
scale invariance, since $L$ transforms under (29)-(32), while $M$ 
is choosen by the boundary conditions.

The variation with respect to $g^{\mu \nu}$ gives

\begin{equation}
\frac{1}{\kappa}R_{\mu \nu} = - \frac {y^{p}}{2m^{4(p-1)}}g_{\mu \nu}
+ \frac{1}{2} \phi_{, \mu} \phi_{,\nu} 
\end{equation}

Since we can solve for $R = g^{\mu \nu}R_{\mu \nu}$ from (34) and insert in 
(33),
we obtain then an equartion which does not contain the scalar curvature, which
is,

\begin{equation}
\frac{2p-1}{pm^{4(p-1)}}y^{p} = V + M  
\end{equation} 

The variation of the action with respect to $A_{\mu \nu \alpha}$ gives
(recall that $ \chi \equiv \frac{\Phi}{\sqrt{-g}}$)

\begin{equation}
\partial_{\mu} (\chi y^{p-1} \varepsilon^{\mu\nu\alpha\beta}) = 0 
\end{equation}  
which means that 

\begin{equation}  
\chi y^{p-1} = \omega m^{4(p-1)}
\end{equation}      

 where $\omega$ is a dimensionless constant. If $p \neq \frac{1}{2} $ and
$\omega \neq 0 $ 
we see from eqs. (29), (30), (31) and (32) that (37) spontaneously 
breaks scale invariance.  

        To get the physical content of the theory, it is convenient to go     
to the Einstein conformal frame where                                         
\begin{equation}                                                              
\overline{g}_{\mu\nu} = \chi g_{\mu\nu}                                       
\end{equation}                                                                
In terms of $\overline{g}_{\mu\nu}$   the equations can be written 
then in the Einstein     
form ($R_{\mu\nu} (\overline{g}_{\alpha\beta})$ =  usual Ricci tensor)        
\begin{equation}                                                              
R_{\mu\nu} (\overline{g}_{\alpha\beta}) - \frac{1}{2}                         
\overline{g}_{\mu\nu}                                                         
R(\overline{g}_{\alpha\beta}) = \frac{\kappa}{2} T^{eff}_{\mu\nu}             
(\phi)                                                                        
\end{equation}
where                                                                         
\begin{equation}                                                              
T^{eff}_{\mu\nu} (\phi) = \phi_{,\mu} \phi_{,\nu} - \frac{1}{2} \overline     
{g}_{\mu\nu} \phi_{,\alpha} \phi_{,\beta} \overline{g}^{\alpha\beta}          
+ \overline{g}_{\mu\nu} V^{(p)}_{eff} (\phi)                                       
\end{equation}                                                                

and the potential $V^{(p)}_{eff} (\phi)$ is given by 
                
\begin{equation}                                                              
        V^{(p)}_{eff}  = \frac{1}{\omega m^{4(1-\frac{1}{p})}} 
(\frac{p}{2p-1})^{2-\frac{1}{p}} (V(\phi) +M)^{2-\frac{1}{p}}        
\end{equation}

        In terms of the metric $\overline{g}^{\alpha\beta}$ ,
and $V^{(p)}_{eff}$ defined above, the equation    
of motion of the scalar                                                       
field $\phi$ takes the standard General - Relativity form                     
\begin{equation}                                                              
\frac{1}{\sqrt{-\overline{g}}} \partial_{\mu} (\overline{g}^{\mu\nu}          
\sqrt{-\overline{g}} \partial_{\nu}                                           
\phi) + V^{(p) \prime}_{eff} (\phi) = O.                                          
\end{equation}

and for the case of interest, $V(\phi) = fe^{\alpha \phi}$
so that 
\begin{equation}                                                              
        V^{(p)}_{eff}  = \frac{1}{\omega m^{4(1-\frac{1}{p})}}                 
(\frac{p}{2p-1})^{2-\frac{1}{p}} (f e^{\alpha \phi} +M)^{2-\frac{1}{p}}                 
\end{equation}

Notice that $\alpha \phi \rightarrow - \infty$,
 \begin{equation}                                                              
        V^{(p)}_{eff} \rightarrow  \frac{1}{\omega m^{4(1-\frac{1}{p})}}                 
(\frac{p}{2p-1})^{2-\frac{1}{p}} (M)^{2-\frac{1}{p}} = constant.                 
\end{equation}
As in our previous example, there is an asymptotically flat region, 
which can be the region where inflation in the early universe took place.

Notice that under a scale transformation, both $M$ and $\omega$ transform
according to 
\begin{equation}
M \rightarrow  e^{-\theta} M, \omega \rightarrow 
e^{ (\frac{1}{p} - 2) \theta} \omega  
\end{equation}

Note, for example, that the asymptotically value (44) is invariant under (45).
(45) transforms one vacuum into another one.

Let us consider now the case $0 < p < \frac {1}{2}$ and $M<0, f<0$. 
Then,
\begin{equation}                                                              
        V^{(p)}_{eff}  = \frac{1}{\omega m^{4(1-\frac{1}{p})}}                 
(\frac{p}{|2p-1|})^{2-\frac{1}{p}} (|f| e^{\alpha \phi} +|M|)^{2-\frac{1}{p}} 
\end{equation}                                              

and as $\alpha \phi \rightarrow \infty$
\begin{equation}
 V^{(p)}_{eff} \rightarrow   C e^{-(\frac{1}{p} -2) \alpha \phi}
\end{equation} 

(C is a constant) and since $0 < p < \frac {1}{2}$, $\frac{1}{p} -2 > 0$ so that 
$V^{(p)}_{eff} \rightarrow 0 $ in this limit. That is, as the magnitude 
of the original potential $V = |f| e^{\alpha \phi}$ goes to infinity,
the effective potential approaches zero without fine tunning.
Since there is also a region of constant, positive value of the potential,
if $\omega > 0$, the 
potential in question can connect, as was the case of the interpolating 
model of section 5, an inflationary phase with a small (and slowly 
decaying in this case)
cosmological term, responsible fo the slowly accelerated universe today.

Here such an effect has been obtained from spontaneous symmetry breaking of
scale invariance. Kaganovich $^{14}$ has studied the possibility of
obtaining quintessence models from small pertubations of the two measure 
model defined in section 1 where the small perturbations explicitly break 
scale invariance.

More details concerning the scale invariant (i.e. where all scale symmetry
breaking is introduced spontaneously rather than explicitly) approach to
quintessence will be given elsewhere $^{15}$. 

For the case $\alpha = 0$ and $p = \frac{1}{2}$ the theory has an infinite
dimensional extra symmetry. To see this notice first that if $\alpha = 0$ 
the term $\int V \Phi d^{4}x = V \int \Phi d^{4}x$, since $V$ is in 
such a case a constant.
 So this term is dynamically irrelevant, it is the integral of a total 
divergence.

The rest of the action is invariant under the infinite dimensional group
of diffeomorphisms in the internal space of the $\varphi_{a}$

\begin{equation}
\varphi_{a} \rightarrow \varphi^{\prime}_{a} (\varphi_{b})
\end{equation}
so that
\begin{equation}
\Phi \rightarrow J \Phi
\end{equation}

where $J$ is the jacobian of the transformation (48).

The internal diffeomorphism (48) has to be performed together with the conformal 
transformation of the metric, while the three index potential remains 
unchanged.

\begin{equation} 
g_{\mu \nu} \rightarrow J g_{\mu \nu}, A_{\mu \nu \alpha} \rightarrow 
A_{\mu \nu \alpha} 
\end{equation}  

Notice that the point $ p = \frac{1}{2} $, where an extra symmetry appears is 
therefore a true critical point. The physics of the model changes drastically
as we go through this point:

For $p > \frac{1}{2}$,
$\alpha \phi \rightarrow \infty$, means $V^{(p)}_{eff} \rightarrow \infty$,
while if $p < \frac{1}{2}$, 
$\alpha \phi \rightarrow \infty$, means $V^{(p)}_{eff} \rightarrow 0$

This type of "square root gauge theory" has been investigated for the case
of a vector potential gauge field theory $^{16,17}$,
i.e. for an $L$ of the form $\sqrt{F^{\mu \nu} F_{\mu \nu}}$, which has 
identical transformation properties under conformal transformations
of the metric as $y^{\frac{1}{2}}$, with $y$ defined as in (26) and
where no transformation of the gauge fields themselves is assumed.
Also these theories have been studied from the point of view of higher
dimensional physics in the modified measure formalism $^{18}$.

\section{Strings and Branes}
In the case of strings, we can replace in the  
Polyakov action, the measure        
$\sqrt{-\gamma}d^{2}x$ (where $\gamma_{ab}$ is the metric              
defined on the world sheet, $\gamma = det (\gamma_{ab})$ and           
$a,b$ indices for the world sheet coordinates) by $\Phi d^{2}x$,       
where  $\Phi = \varepsilon^{ab} \varepsilon_{ij}                       
\partial_{a} \varphi_{i}\partial_{b} \varphi_{j}$. Then for the        
bosonic string $^{19}$, we consider the action                             
                                                                              
\begin{equation}                                                              
S = - \int d\tau d\sigma \Phi[                                                
\gamma^{ab} \partial_{a} X^{\mu}\partial_{b} X^{\nu} g_{\mu \nu}              
- \frac {\varepsilon^{ab}}{\sqrt{-\gamma}} F_{ab} ]                           
\nonumber                                                                     
\end{equation}

where $ F_{ab} = \partial_{a} A_{b} - \partial_{b} A_{a} $ and                
$A_{a}$ is a gauge field defined in the world sheet of the string.            
The term with the gauge fields is irrelevant if the ordinary                  
measure of integration is used, since in that case it would be a divergence,  
but is needed for a consistent dynamics in the modified measure reformulation 
of string theory. This is due to the fact that if we avoid such a             
contribution to the action, one can see that the variation of the             
action with respect to $\gamma^{ab}$ leads                                    
to the vanishing of the induced metric on the string.                         
The equation of motion obtained from the variation of the                     
gauge field $A_{a}$ is                                                        
$\varepsilon^{ab}\partial_{a} (\frac{\Phi}{\sqrt{-\gamma}}) = 0$ . From which 
we obtain that $\Phi = c \sqrt{-\gamma}$ where $c$ is a constant which can be 
seen is the string tension. The string tension appears then as an integration 
constant and does not have to be introduced from the beginning. The string    
theory Lagrangian in the modified measure formalism does not have any         
fundamental scale associated with it. The gauge field strength $ F_{ab}$ can  
be solved from a fundamental constraint of the theory, which is obtained      
from the variation of the action with respect to the measure fields           
$\varphi_{j}$ and which requires that $L=M=constant$. Consistency demands     
$M=0$ and finally all the equations are the same as those of standard         
bosonic string theory.

Extensions to both the super symmetric case $^{20}$ and to  higher branes        
are possible $^{19, 20}$. When considering the modified measure reformulation of  
the super string, it is very useful to consider the Siegel reformulation of   
the Green Schwarz action, where the Wess-Zumino term is the square of         
super symmetric currents $^{21}$. Then the modified measure action will be given  
by $S = \int d^{2}x \Phi L$, where $L$ is given by                            
\begin{equation}                                                              
L= \frac{1}{2} \gamma^{ab} J^{\mu}_{a} J_{\mu b}                              
-i\frac {\varepsilon^{a b}}{\sqrt{-\gamma}} J^{\alpha}_{a} J_{\alpha b}       
\nonumber                                                                     
\end{equation}                                                                
Here the super symmetric currents are defined by                              
 $J^{\alpha}_{a} = \partial_{a} \theta^{\alpha},                              
J^{\mu}_{a} = \partial_{a} X^{\mu} -                                          
 i \partial_{a} \theta^{\alpha}\Gamma^{\mu}_{\alpha \beta} \theta^{\beta}$    
and finally the current $J_{\alpha a}$, that contains all the dependence      
on the compensating fields $\phi_{\alpha}$ introduced by Siegel $^{21}$           
to achieve super symmetry invariance (and not just super symmetry up to       
a total divergence of the usual formulation), is defined as
\begin{equation}                                                              
J_{\alpha a} = \partial_{a} \phi_{\alpha} -                                   
2i(\partial_{a} X^{\mu})\Gamma_{\mu \alpha \beta} \theta^{\beta}              
- \frac{2}{3} (\partial_{a} \theta^{\beta}) \Gamma^{\mu}_{ \beta \delta}      
\theta^{\delta}\Gamma_{\mu \alpha \epsilon} \theta^{\epsilon}                 
\nonumber                                                                     
\end{equation}                                                                
                                                                              
 Then, as opposed to the Siegel case, due to the use of the modified measure, 
the compensating fields $\phi_{\alpha}$, do not enter in the action as        
in a total divergence (that is they are dynamically irrelevant). Instead,     
these compensating fields are responsible for the existence of the gauge      
field $A_{a}$, which we  explained for the case of the bosonic string         
and which does not have to be introduced independently of the Siegel          
compensating fields and in fact it can be read of from the above action to be
$A_{a} = i\theta^{\alpha} \partial_{a} \phi_{\alpha}$.                        
 The gauge field needed for consistent dynamics is now                        
a composite structure of a nature reminiscent to those considered in Ref.22.

For the case of higher p-branes, in the bosonic case, a term of the form      
$\frac {\varepsilon^{a_{1}a_{2}...a_{p+1}}}{\sqrt{-\gamma}}                   
\partial_{[a_{1}}A_{a_{2}...a_{p+1}]}$ has to be considered instead of        
the $\frac {\varepsilon^{ab}}{\sqrt{-\gamma}} F_{ab} $ considered for the     
string. As in the case of strings, the variation with respect to the          
$\varphi_{j}$ fields requires $L=M=constant$. For higher branes however,      
a consistent dynamics is achieved as long as $M \neq 0$. Once this is done,   
in contrast to the standard formulation of Polyakov-type actions for branes,  
no explicit cosmological term in the brane has to be added. It is interesting 
to recall at this point that the original motivation $^{1}$ for the use of a     
modified measure was in the context of the cosmological constant problem in   
field theoretical approaches to                                               
gravitation and in the context of the theory of extended objects the          
modified measure approach continues to be useful concerning how to handle     
(which in this case means avoiding) the cosmological constant defined in the  
world brane.                                                                  
Super branes can also be formulated with the use of a modified measure.       
In this case the Bergshoeff-Sezgin formalism $^{23}$, which generalizes for       
the case of higher branes the Siegel formalism $^{21}$ 
has to be used. If we      
do this, we find again that the field $A_{a_{1}...a_{p}}$, which is required  
for a consistent dynamics in the modified measure formalism, does not have    
to be introduced separately, instead it appears as a consequence of the       
Bergshoeff-Sezgin formalism $^{20}$. Again, the Bergshoeff-Sezgin 
compensating    
fields are dynamically relevant, unlike the situation in the standard         
case $^{23}$. Furthermore, in contrast to the treatment of Ref.$23$, 
no explicit cosmological        
term in the brane needs to be included. As it was explained in the            
string case, the brane tension appears as an integration constant of the      
equations of motion. Once again, no fundamental scales need to be             
introduced in the original action of the theory.                              

\section{Discussion and Conclusions}
Here we have seen that the consideration of a volume element independent of
the metric allows i) to handle the cosmological constant problem, ii) to
produce new realistic gravitational theories which allow spontaneous 
symmetry breaking (ssb) of scale invariance iii) that this ssb of scale
invariance does not necessarily requires the existence of Goldstone bosons 
and iv) that string and brane theories without a fundamental scale are 
possible.

Concerning the gravitational theory, one should notice that it was crucial
to get the physical content of the theory to go to the metric 
$\overline{g}_{\mu\nu} = \chi g_{\mu\nu}$, where 
the theory takes the Einstein form.
 
It has been the subject of great debate which conformal frame is the physical
one. For a review, see Ref.$24$.

In our case it appears that $ \overline{g}_{\mu\nu} = \chi g_{\mu\nu}$ is
indeed the correct choice.

This is more transparent if we look at the theory in the Hamiltonian 
formalism. In Hamiltonian languge the quantization of the theory and the
statistical mechanichs phase space are more directly available.

In this case it is immediate to see that the original metric $ g_{\mu\nu}$
has vanishing canonically conjugate momenta. In contrast to this, it is
a simple matter to see that all the canonically conjugate momenta to the
connections $\Gamma^{\lambda}_{\mu \nu}$ are functions of 
$\overline{g}_{\mu\nu} = \chi g_{\mu\nu}$  .
 So that $\Gamma^{\lambda}_{\mu \nu}$ and
$ \overline{g}_{\mu\nu} $ are conjugate variables, while the original metric
does not have a conjugate variable.

Furthermore, very much connected to this, in the quantum theory, 
it is the expansion of 
$ \overline{g}_{\mu\nu}$ in creation and annihilation operators (after 
suitable gauge fixing) the one that provides us with the correctly 
normalized graviton states. In contrast the original metric has unphysical
singularities for $V+M=0$ in the two measure model for example, while no
such a problem appears in the Einstein frame. This shouldn't worry us, in 
phase space we don't even see the original metric.

Another issue for further study is the connection of the modified measure
formalism to of brane scenarios. It appears for example that the
two measure formalism may be related to such type of scenario, which 
automatically assigns a different measure for the bulk and for 
the brane $^{25}$.

Furthermore, the quintessence picture (section 6) presents some remarkable 
points of contact with what is subsequently discussed in connection to
the modified measure formulation of brane theories in that both require
the use of antisymmetric potentialsfor achieving consistent dynamics. 
This point also deserves
further research.

Another direction which seems to have a more direct connection to what is done 
here is the discrete Kaluza Klein approach developed in Ref.$26$ , where
two copies of space-time are introduced in order to develop a non commutative
geometry. In a contex like this the measure $\Phi$ can naturally arise as the
jacobian of the mapping of these two spaces.

Finally let us point out that many of the results obtained here agree with the
considerations of Bekenstein and Meisels $^{27}$, which were considering
local rather than global scale invariance. In any case, they found that local
scale invariance goes together with a constant vacuum energy and constant
particle masses, exactly what it is found here in the Einstein frame in the
unbroken sector of the theory. For example when $M$ can be ignored in the two measure
theory. In the quintessence scale invariant model the analysis is more 
complex because there are now two sources of scale symmetry breaking ($M$
and $\omega$) . This will be studied further in the future $^{15}$.   
\section{Acknowledgements}
I would like to thank J.Alfaro, M. Banados, J.Bekenstein, N.Berkovits, Z.Bern, 
L. Cabral, S. del Campo, C.Castro,  F. Cooperstock, 
A.Davidson, M.Dine, P.Gaete,M.Giovannini, F.Hehl, 
L.Horwitz, A.Kaganovich, M.Loewe, Y.Ne'eman, M.Pavsic, J.Portnoy, 
E.Spallucci, J.Schwarz, Y.Verbin and K.Wali for conversations
on the subjects of this paper.


\begin{thebibliography}{99}                                                   
\bibitem{modmes} For a review of theories of this type and further references 
see,                                                                          
E.I.Guendelman and A.B.Kaganovich,                                            
Phys. Rev. D 60, 065004 (1999) and for the connection of these theories       
to  theories of composite gauge fields and volume preserving diffeomorphisms 
see E.I.Guendelman Int. Journ. Mod. Phys. A14, 3497 (1999).                       
                                                                         
\bibitem{decl} E.I.Guendelman,                                                
Mod. Phys. Lett. A14, 1043 (1999).                                            
                                                                              
\bibitem{cosco} E.I.Guendelman,                                               
Mod. Phys. Lett. A14, 1397 (1999);  E.I.Guendelman, gr-qc/9901067;            
E.I.Guendelman, "Scale Invariance and Cosmology", in Proc. of
the 8-th Canadian Conference                                                  
on General Relativity and Relativistic Astrophysics, Montreal, pp. 201 (1999);
E.I.Guendelman, hep-th/0008122  (contribution to ICHEP 2000, Osaka, Japan,
27 Jul- 2 Aug 2000) ; E.I.Guendelman, gr-qc/0004011 (contribution to the 35th
Rencontres de Moriond: Energy Densities in the Universe, Les Arcs, Savoie, 
France, 22-29 Jan 2000).
                                                                            
\bibitem{gold} E.I.Guendelman,                                                
Class. Quantum Grav. 17, 261 (2000).                                          
                                                                              
\bibitem{Zee} A.Zee,                                                          
Phys. Rev. Lett. 42, 417 (1979).                                              
                                                                              
\bibitem{K-i} K-i. Maeda, J.A.Stein-Schabes                                   
and T.Futamase, Phys. Rev. D39, 2848 (1989).                                  
                                                                              
\bibitem{infla} F.S.Accetta, D.J.Zoller and M.S.Turner,                       
 Phys. Rev. D31, 3046 (1985); F.Lucchin, S.Matarrese                          
and M.D.Pollock,  Phys. Lett. 167B, 163 (1986);                               
 R.Fakir and W.G.Unruh, Phys.Rev.D41, 1792 (1990);                            
J.L.Cervantes-Cota and H.Dehnen,  Phys. Rev.D51, 395 (1995);                  
D.I.Kaiser,Phys. Rev.D49, 6347 (1994) and Phys. Rev.D52,4295 (1995);          
E.Komatsu and Futamase, Phys. Rev.D58, 023004 (1998); E.Komatsu               
and Futamase, Phys. Rev.D59, 064029 (1999).                                   

\bibitem{sta} A.A.Starobinskii, Astron. Lett. 9, 302 (1983).   
\bibitem{barrow} J.D.Barrow and A.C.Otterwill, 
J.Phys. A: Math. Gen. 16, 2757 (1983); J.D.Barrow and S.Cotsakis, 
Phys. Lett. B214, 515 (1988); K-i. Maeda Phys. Rev. D37, 858 (1988). 
\bibitem{mijic} M.B.Mijic, M.M.Morris and W-M Suen, 
Phys. Rev. D34, 2934 (1986).                                                                              
\bibitem{seesaw} For a review see B.Kayser, with F.Gibrat-Debu and F.Perrier, 
The Physics of Massive Neutrinos, World Scientific, 1989.         
\bibitem{larry} For a review see L.Krauss, Sci. Amer. 280, 34 (1999).            
\bibitem{quint} Some references on this subject are: C.Wetterich, Nucl. Phys.
B302, 668 (1988); B.Ratra and P.J.E.Peebles, Phys. Rev. D37, 3406 (1988);
P.J.E.Peebles and B.Ratra, Astrophys. J. 325, L17 (1988); R.Caldwell, R.Dave
and P.Steinhardt, Phys. Rev. Lett. 80, 1582 (1998); N.Weiss, Phys. Lett. B197,
42 (1987); Y.Fujii and T.Nishioka, Phys. Rev. D42, 361 (1990); M.S.Turner
and M.White, Phys. Rev. D56, R4439 (1997); E.Copeland, A.Liddle and D.Wands,
Phys. Rev. D57, 4686 (1998); C.T.Hills, D.N.Schramm and J.N.Fry, Comments
Nucl.Part. Phys. 19, 25 (1989); J.Frieman, C.Hill and R.Watkins,
Phys. Rev. D46, 1226 (1992); J.Frieman, C.Hill, A.Stebbins and I.Waga, 
Phys. Rev. Lett. 75, 2077 (1995); C.Wetterich, Nucl. Phys. B302, 645 (1988);
Astron. Astrophys. 301, 321(1995); P.Ferreira and M.Joyce, Phys. Rev. Lett. 
79, 4740 (1997); Phys. Rev. D58, 023503 (1998); I.Zlavtev, L.Wang and 
P.Steinhardt, Phys. Rev. Lett. 82, 896 (1999); P.Steinhardt, L.Wang and
I.Zlatev, Phys. Rev. D59, 123504 (1999); P.J.E.Peebles and A.Vilenkin, Phys. 
Rev. D59, 063505 (1999).
\bibitem{kaga} A.B.Kaganovich,  hep-th/0007144.
\bibitem{guend} E.I. Guendelman, manuscript in preparation.
\bibitem{aurilia} A.Aurilia, A.Smailagic and E.Spallucci,
Phys. Rev. D47, 2536 (1993).                                                             
\bibitem{amer}  N. Amer and E.I.Guendelman, to appear in Int. Journ. 
Mod. Phys. A (2000).                                                              
\bibitem{fiveD} E.I.Guendelman, Phys. Lett. B412, 42 (1997).                                                              
\bibitem{string}E.I. Guendelman, Class. Quant. Grav. 17, 3673 (2000).
\bibitem{super}E.I. Guendelman, hep-th/0006079.
\bibitem{siegel}W.Siegel, Phys. Rev. D50, 2799 (1994). 
\bibitem{comp} E.Nissinov, E.I. Guendelman and S.Pacheva, Phys. Lett. B360,
57 (1995); hep-th/9903245; C.Castro, Int.J.Mod.Phys. A13, 1263 (1998). 
\bibitem{bergs} Bergshoeff and E.Sezgin, Phys. Lett B356 (1995).
\bibitem{conf} V.Faraoni, E.Gunzig and P.Nardone, gr-qc/9811047.
\bibitem{brane} N.Arkani-Hamed, S.Dimopoulos, G.Dvali and N.Kaloper,
Phys. Rev. Lett. 84, 586 (2000).
\bibitem{dkk} N.A.Viet and K.Wali, Int. J. Mod. Phys. A11, 2403 (1996).
\bibitem{beken} J.D.Bekenstein and A.Meisels, Phys. Rev. D22, 1313 (1980). 

\end{thebibliography}
\end{document}